\documentclass[conference]{IEEEtran}
\IEEEoverridecommandlockouts

\usepackage{cite}
\usepackage{amsmath,amssymb,amsfonts}
\usepackage{algorithmic}
\usepackage{graphicx}
\usepackage{textcomp}
\usepackage{xcolor}
\usepackage{subcaption}
\usepackage{hyperref}

\newtheorem{theorem}{Theorem}
\newtheorem{definition}{Definition}

\newtheorem{lemma}{Lemma}
\newtheorem{corollary}{Corollary}
\newtheorem{remark}{Remark}

\newcommand{\remove}[1]{}

\allowdisplaybreaks
\usepackage{etoolbox}
\makeatletter
\patchcmd{\@begintheorem}{\textit}{\textbf}{}{}
\let\sv@thm\@thm
\def\@thm{\let\indent\relax\sv@thm}
\makeatother

\def\BibTeX{{\rm B\kern-.05em{\sc i\kern-.025em b}\kern-.08em
    T\kern-.1667em\lower.7ex\hbox{E}\kern-.125emX}}
\begin{document}

\title{A Finite-Sample Strong Converse for Binary Hypothesis Testing via (Reverse) R\'enyi Divergence\\
}

\author{\IEEEauthorblockN{Roberto Bruno}
\IEEEauthorblockA{
\textit{University of Salerno}\\
 Fisciano (SA), Italy \\
rbruno@unisa.it}
\and
\IEEEauthorblockN{Adrien Vandenbroucque}
\IEEEauthorblockA{
\textit{EPFL}\\
Lausanne, Switzerland \\
adrien.vandenbroucque@epfl.ch}
\and
\IEEEauthorblockN{Amedeo Roberto Esposito}
\IEEEauthorblockA{
\textit{Okinawa Institute of Science and Technology}\\
Onna, Japan \\
amedeo.esposito@oist.jp}
}

\maketitle

\begin{abstract}
This work investigates binary hypothesis testing between $H_0\sim P_0$ and $H_1\sim P_1$ in the finite-sample regime under asymmetric error constraints. By employing the ``reverse" R\'enyi divergence, we derive novel non-asymptotic bounds on the Type II error probability which naturally establish a strong converse result. Furthermore, when the Type I error is constrained to decay exponentially with a rate $c$, we show that the Type II error converges to 1 exponentially fast if $c$ exceeds the Kullback-Leibler divergence $D(P_1\|P_0)$, and vanishes exponentially fast if $c$ is smaller. Finally, we present numerical examples demonstrating that the proposed converse bounds strictly improve upon existing finite-sample results in the literature.
\end{abstract}

\begin{IEEEkeywords}
Hypothesis Testing, Finite Samples, R\'enyi Divergence, Data-Processing Inequality, Strong converse
\end{IEEEkeywords}

\section{Introduction}
Binary hypothesis testing constitutes a fundamental problem in statistical inference. 
In the standard setting, an observer collects $n$ independent and identically distributed (i.i.d.) samples from an unknown probability measure $P$. The objective is to decide between two hypotheses: the null hypothesis ($H_0$), which posits that $P=P_0$, and the alternative hypothesis ($H_1$), which posits that $P=P_1$. Binary hypothesis testing finds widespread application in various domains, ranging from signal processing and communications \cite{kay1993fundamentals, van2004detection} to cybersecurity \cite{chandola2009anomaly, patcha2007overview} and machine learning \cite{bishop2006pattern, hastie2009}, to name a few.

\smallskip
In binary hypothesis testing, depending on the decision, two types of error may occur: the Type I error (or false alarm) which consists of incorrectly rejecting the null hypothesis $H_0$ when it is true; and the Type II error (or missed detection), which corresponds to failing to reject $H_0$ when the alternative hypothesis $H_1$ holds.  
Ideally, one aims to minimize both probabilities of error; however, a fundamental trade-off exists between these two quantities, preventing the arbitrary reduction of one error probability without affecting the other, and thus making their independent simultaneous minimization unfeasible in many cases. To address this, a common approach that corresponds to the classical Neyman-Pearson formulation is to minimize the Type II error probability subject to a constraint on the Type I error probability. In this setting, the Neyman-Pearson lemma \cite{neyman1933ix} identifies the Log-Likelihood Ratio Test as the optimal decision rule, achieving the best trade-off possible among the two errors for any sample size. 

In the asymptotic regime, as the number of samples $n$ tends to infinity, the fundamental limit of the trade-off between the two probabilities of error is characterized by the Chernoff-Stein lemma \cite{chernoff1956large, chernoff1952measure,cover1999elements}, which establishes that the optimal error exponent for the Type II error is the Kullback-Leibler (KL) divergence, $D(P_0\| P_1)$, {when the Type I error is bounded by a constant. This result has been successively refined and extended to the finite blocklength regime by Espinosa \textit{et al.}  \cite{Espinosa}, who showed that this optimal rate holds even when the Type I error is not bounded by a constant but decays sub-exponentially with $n$. Conversely, when the Type I error is constrained to decay exponentially with a rate $c$, Blahut \cite{Blahut} as well as Han and Kobayashi \cite{Han_strongconverse} identified $D(P_1\|P_0)$ as the maximal achievable rate under which the Type II error can asymptotically vanish.}

Although these results provide crucial information about the problem structure, the majority only provide asymptotic guarantees. 
For this reason, recent research has increasingly shifted focus toward the finite samples regime, employing various information-theoretic measures to derive converse and achievability bounds \cite{polyanskiy2010channel, bar2002complexity,pensia2024sample,lungu2024finite}.

Among these measures, the Total Variation ($\mathrm{TV}$) distance and the $E_{\gamma}$ divergence provide exact characterizations of the error trade-off. Specifically, for an optimal test the sum of the two errors relates to the TV distance as follows \cite{lehmann2005testing}
\begin{equation*}\label{eq:1-tv}
    \min_{\phi} \{P_0^{n}(\phi(\mathbf{X})=1) + P_1^{n}(\phi(\mathbf{X})=0)\} = 1 - \mathrm{TV}(P_0^{n},P_1^{n}).
\end{equation*}
A similar characterization of the error can also be established for the $E_{\gamma}$ divergence, which represents a natural generalization of the $\mathrm{TV}$ distance \cite{asoodeh2020contraction, mullhaupt2025bounding}. However, both $\mathrm{TV}$ and $E_{\gamma}$ suffer from a critical limitation: they do not tensorize. As a result, the corresponding bounds for product distributions become computationally intractable as the sample size $n$ increases, as shown by Bhattacharyya \textit{et al.} \cite{bhattacharyya2022approximating, bhattacharyya2025total}.

Thus, it is natural to focus on information measures that tensorize, as they lead to tractable and scalable bounds.
Polyanskiy \textit{et al.} \cite{polyanskiy2010channel} derived finite-length bounds on the Type II error probability, with the KL divergence between $P_0$ and $P_1$ appearing as the dominant term. Subsequently, building on a similar approach, Lungu and Kontoyiannis \cite{lungu2024optimal} further sharpened the results by providing tighter finite-sample bounds.
However, these KL-based bounds (often derived via Fano-type inequalities) yield weak converse results: 
they can only establish that when the Type I error constraint is too stringent, the Type II error probability cannot vanish (i.e., it is bounded away from 0). A strong converse, on the other hand, strengthens the statement showing that under such conditions the Type II error probability converges to one.  Achieving strong converse results generally requires additional techniques, such as the blowing-up lemma~\cite{ahlswede1976bounds} or the smoothing-out method~\cite{liu2020second}. Without these refinements, such bounds fail to capture the 
sharp phase transition that separates the achievable and impossible regimes.

Along a different line, Bar-Yossef \cite{bar2002complexity}
established bounds on the Type II error probability in terms of the Squared Hellinger distance between $P_0$ and $P_1$ and the sample size $n$.\remove{Similarly, Pensia \textit{et al.} \cite{pensia2024sample} 
focused on the Bayesian setting, where the hypotheses are endowed with prior probabilities and the objective is to minimize the average error probability, and provided a rigorous characterization of the sample complexity---the minimum number of observations required to meet specific error constraints---in terms of Hellinger distances as well.
However, these Hellinger-based approaches are inherently symmetric and therefore do not capture the asymmetric nature of binary hypothesis testing. Moreover, as with KL-based approaches, the resulting bounds yield only weak converse guarantees.}
Moreover, Pensia \textit{et al.} \cite{pensia2024sample,kazemi2025sample} showed that the asymmetric and Bayesian settings, where the hypotheses are endowed with prior probabilities and the objective is to minimize the average error probability, can be treated equivalently. Thus, focusing on the Bayesian setting, they provided a rigorous characterization of the sample complexity---the minimum number of observations required to meet specific error constraints---in terms of Hellinger/Jensen-Shannon divergences. 

We focus on characterizing strong converse behaviour in finite sample regimes, and capturing the phase transition between the achievable and impossible regimes. In pursuing this goal, our analysis led us to the R\'enyi divergence \cite{vanerven_renyi}, which has previously been shown to be closely related to hypothesis testing and strong converse phenomena  \cite{mosonyi2015quantum, polyanskiy2010arimoto}. 

\subsection{Contributions}
In this work, we fully characterize the phase transition of the Type II error probability in binary hypothesis testing when the Type I error is constrained to decay exponentially. Specifically, in Section \ref{sec:phase_transition}, assuming that the Type I error probability decays as $e^{-nc}$, we demonstrate that the Type II error probability converges to 1 exponentially fast if $c>D(P_1\|P_0)$, and vanishes if $c<D(P_1\| P_0)$ as the sample size $n$ grows. This result provides a quantitative finite-sample counterpart to the asymptotic one presented in \cite{Blahut,Han_strongconverse}.

To establish the phase transition, in Section \ref{sec:bounds}, we derive novel non-asymptotic lower bounds on the Type II error using the R\'enyi divergence of order $\lambda>1$, alongside an upper bound expressed in terms of the R\'enyi divergence of order $\lambda\in(0,1)$. These bounds naturally yield a strong converse result without relying on additional auxiliary refinements. As a result, our approach differs fundamentally from classical strong converse techniques\cite{ahlswede1976bounds,liu2020second,tyagi2019strong,oohama2018exponential}.

Furthermore, a key aspect of our approach relies on the use of the ``reverse" R\'enyi divergence, $D_{\lambda}(P_1\| P_0)$, rather than the more common ``forward" direction from $P_0$ to $P_1$. This also fundamentally distinguishes our approach from those based on symmetric measures, such as the $\mathrm{TV}$ distance and the Squared Hellinger distance.

Finally, in Section~\ref{sec:plots}, we conclude by showing through numerical examples that our converse bound strictly improves upon existing finite-sample bounds in the literature in both discrete and continuous settings and under various Type I error regimes.

\section{Preliminaries and notation}

Let $\mathbf{X} = (X_1, X_2, \ldots, X_n)$ denote a vector of $n$ observed i.i.d. random variables with each $X_i \sim P$ distributed according to a probability measure $P$ on the measurable space $(\mathcal{X}, \mathcal{B})$. In the context of binary hypothesis testing, we assume that the observations are generated according to one of two hypotheses:

\begin{itemize}
    \item \textbf{Null Hypothesis ($H_0$):} Under this hypothesis, $P=P_0$. Consequently, the joint probability distribution of the random vector $\mathbf{X}$ is given by the product measure $P_0^{n}$ on the measurable space $(\mathcal{X}^n, \mathcal{B}^n)$.
    
    \item \textbf{Alternative Hypothesis ($H_1$):} Under this hypothesis,  $P = P_1$. Consequently, the joint probability distribution of the random vector $\mathbf{X}$ is given by the product measure $P_1^{n}$ on the same measurable space $(\mathcal{X}^n, \mathcal{B}^n)$.
\end{itemize}
We denote by  $p_0(\mathbf{x})$ (resp. $p_1(\mathbf{x})$) the probability density function of $\mathbf{X}$ under $H_0$ (resp. $H_1$) which is the Radon-Nikodym derivative of $P_0^{n}$ (resp. $P_1^{n}$) with respect to a dominating measure $\mu$. 

The fundamental objective is to design a decision rule, or \emph{test}, that given $\mathbf{X}$ decides which of the two hypotheses is true, i.e., if $\mathbf{X}^n \sim P_0^{n}$ or $\mathbf{X}^n \sim P_1^{n}$.
More formally, the decision rule can be described as a binary function $\phi: \mathcal{X}^n \to \{0, 1\}$ such that:

\begin{equation}
    \phi(\mathbf{x}) = \begin{cases}
         1 &\text{if } \mathbf{x} \in A, \\
         0 &\text{if } \mathbf{x} \in A^c,
    \end{cases}
\end{equation}
where $A \subset \mathcal{X}^n$ is the region in which we accept the alternative hypothesis ($H_1$), and $A^c = \mathcal{X}^n \setminus A$ is the region in which we reject the alternative hypothesis, i.e., we accept the null hypothesis ($H_0$).

Depending on the decision, two possible types of error may occur:
\begin{itemize}
    \item \textbf{Type I Error (False Alarm):} The probability of incorrectly rejecting $H_0$ when $H_0$ is true. This probability is conventionally denoted by $\alpha_n$:
    \begin{align*}
        \alpha _n &= \mathbb{P}(\text{Decide } H_1 \mid H_0) = P_0^{n}(\phi(\mathbf{X})=1)\\ &= P_0^{n}(A) = \int_{A} p_0(\mathbf{x}) \, d\mu(\mathbf{x}).
    \end{align*}

    \item \textbf{Type II Error (Missed Detection):} The probability of incorrectly accepting $H_0$ when $H_1$ is true. This probability is conventionally denoted by $\beta_n$:
    \begin{align*}
        \beta_n &= \mathbb{P}(\text{Decide } H_0 \mid H_1) = P_1^{n}(\phi(\mathbf{X})=0)\\ &= P_1^{n}(A^c) = \int_{A^c} p_1(\mathbf{x}) \, d\mu(\mathbf{x}).
    \end{align*}
\end{itemize}

In the following, we focus on the asymmetric setting: our objective is to minimize the Type II error probability while ensuring that the Type I error probability remains below a specified threshold $\varepsilon\in(0,1)$. To this end,  let $\beta_n(\varepsilon)$ denote the minimum Type II error probability achievable by any decision rule using $n$ samples, subject to the constraint that the Type I error probability does not exceed $\varepsilon\in (0,1)$, i.e., $\alpha_n\leq \varepsilon$. More formally:

\begin{equation}\label{eq:Beta_eps_def}
    \beta_n(\varepsilon)=\inf_\phi \big\{{P_1^{n}(\phi(\mathbf{X})=0}) : P_0^{n}(\phi(\mathbf{X})=1)\leq \varepsilon\big\}.
\end{equation}
Note that $\varepsilon$ can also be chosen to decay as a function of the sample size $n$.

The Neyman-Pearson lemma \cite{neyman1933ix} states that the optimal solution to \eqref{eq:Beta_eps_def} is achieved by a Log-Likelihood Ratio Test (LLRT). We recall that a deterministic LLRT with threshold $\tau\in \mathbb{R}$ is a decision rule $\phi:\mathcal{X}^n \to \{0,1\}$ such that
\begin{equation}
    \phi(\mathbf{x)}=\begin{cases} 1 &\quad \text{if } \log\frac{P_1^{n}(\mathbf{x})}{P_0^{n}(\mathbf{x})} \geq \tau,\\  0 &\quad \text{otherwise.} \end{cases}
\end{equation}
In other words, the test accepts the alternative hypothesis $H_1$ if the log-likelihood ratio is greater than or equal to the threshold $\tau$, and rejects it otherwise.

Throughout the paper, we use $\log$ to denote the natural logarithm.

\subsection{R\'enyi Divergence}

Let $P$ and $Q$ be two probability measures in the measurable space $(\mathcal{X}, \mathcal{B})$ and let $p(\mathbf{x})$ and $q(\mathbf{x})$ be their densities.

\begin{definition}
For $\lambda \in (0, \infty) \setminus \{1\}$, the R\'enyi divergence of order $\lambda$ of $P$ from $Q$ is defined as follows~\cite{vanerven_renyi}:
\begin{equation}
  D_{\lambda}(P \| Q) =  \frac{1}{\lambda - 1} \log  \int_{\mathcal{X}} p(\mathbf{x})^{\lambda} q(\mathbf{x})^{1-\lambda} \, d\mu(\mathbf{x}).
\end{equation}
When $\lambda \to 1$, one recovers the KL divergence.
\end{definition} 

\section{Error characterization via R\'enyi}\label{sec:bounds}
\subsection{Converse Bounds}\label{sec:converse}

In this section, we derive lower bounds on $\beta_n(\varepsilon)$ by employing the R\'enyi divergence of order $\lambda>1$. Specifically, in Theorem \ref{theorem:beta_lower_bound_renyi_additive}, whose proof is deferred to Appendix \ref{app:thm_1}, we present bounds in terms of both the usual ``forward" divergence, $D_\lambda(P_0^{n}\| P_1^{n})$, and the ``reverse" divergence, $D_\lambda(P_1^{n} \| P_0^{n})$. 

\begin{theorem}\label{theorem:beta_lower_bound_renyi_additive}
    Let $P_1^{n}$ and $P_0^{n}$ be mutually absolutely continuous. For any $\varepsilon \in(0,1)$ it holds that
    
     \begin{align}\label{eq:beta_lower_bound_renyi_additive}
       \beta_n(\varepsilon)\geq&  \max\Bigg\{1-\inf_{\lambda>1}(\varepsilon\, e^{nD_\lambda(P_1\| P_0)})^{\frac{\lambda-1}{\lambda}}, \nonumber\\&\quad\sup_{\lambda>1}\left((1-\varepsilon)^{\frac{\lambda}{\lambda-1}}\,e^{-nD_\lambda(P_0\| P_1)}\right)\Bigg\}.
   \end{align}
\end{theorem}

The bound in \eqref{eq:beta_lower_bound_renyi_additive} consists of two distinct lower bounds. The first one, derived from the ``reverse'' divergence $D_\lambda(P_1^{n} \| P_0^{n})$, can be rewritten as 
\begin{equation*}
    \beta_n(\varepsilon)\geq1-\inf_{\lambda>1}( e^{n(D_\lambda(P_1\| P_0)- \frac{\log(1/\varepsilon)}{n})})^{\frac{\lambda-1}{\lambda}}.
\end{equation*}
This representation makes explicit the dependence on the Type I error constraint. In particular, if there exists some $\lambda>1$ such that $\log(1/\varepsilon)>nD_{\lambda}(P_1\|P_0)$, then the exponent of the subtracted term is negative and thereby it vanishes as $n$ increases, yielding $\beta_n(\varepsilon)\to 1$. Thus, this bound is especially effective in regimes where the Type I error probability decays as a function of $n$. For this reason, it is instrumental in deriving the strong converse result in Section \ref{sec:phase_transition}.

In contrast, the second bound, stemming from the ``forward" divergence, $D_\lambda(P_0^{n}\| P_1^{n})$, ensures  a non-trivial bound even in settings where the first one may become loose ( i.e., negative). For instance, when $\varepsilon$ is fixed and the sample size $n\to \infty$, the inequality $\log(1/\varepsilon)<nD{\lambda}(P_1\|P_0)$ holds for any $\lambda>1$. As a result, in the first bound the subtracted term grows with $n$, rendering the bound vacuous for large $n$. In contrast, for a fixed $\varepsilon$, the second bound yields a lower bound on $\beta_n(\varepsilon)$ of the form $e^{-nD_\lambda(P_0\|P_1)}$.

Furthermore, since the previous lower bounds are non-asymptotic, they can be used to  characterize the finite sample complexity, as shown in the following corollary, whose proof is deferred to Appendix \ref{app:cor}.

\begin{corollary}\label{cor:1}
Let $\lambda>1$ and $\varepsilon\in(0,1)$. Then, for any $\delta\in(0,1)$, a necessary condition to ensure $\beta_n(\varepsilon)\leq \delta$ is that the sample size must satisfy
\begin{align}\label{eq:sample_lb}
     n\geq& \max\Bigg\{\frac{1}{D_{\lambda}(P_0\| P_1)}\left(\log \frac{1}{\delta}-\frac{\lambda}{\lambda-1}\log\frac{1}{1-\varepsilon}\right),\nonumber\\
     &\quad\frac{1}{D_\lambda(P_1\|P_0)}\left(\log\frac{1}{\varepsilon}-\frac{\lambda}{\lambda-1}\log\frac{1}{1-\delta}\right)\Bigg\}.
\end{align}
\end{corollary}
\begin{remark}
    A lower bound characterizing the sample complexity in the Bayesian setting appears in~\cite[Thm. 2]{kazemi2025sample}. By using the equivalence between the Bayesian and asymmetric settings noted in~\cite[Claim 4.6]{pensia2024sample}, it can be rewritten in our notation as
    \begin{equation}
        n\geq \frac{1}{2}\frac{\lambda_*}{1-\lambda_*}\frac{\log(1/2\varepsilon)}{D_{\lambda_*}(P_0\|P_1)},\label{eq:pensia_bound}
    \end{equation}
    where $\lambda_*=\frac{\log(1/2\delta)}{\log(1/2\delta)+\log(1/2\varepsilon)}\in[0.5, 1)$. The range of $\lambda$ in Eq. ~\eqref{eq:pensia_bound} is different from the one in Eq.~\eqref{eq:sample_lb}. A direct comparison is thus not straightforward: in some regimes, Eq.~\eqref{eq:pensia_bound} yields a tighter bound than Eq.~\eqref{eq:sample_lb}, while in others the reverse holds.
\end{remark}

\subsection{Achievability Bound}\label{sec:achievability}
In this section, we shift our focus from converse bounds to achievability results. Specifically, we analyze the performance of an arbitrary LLRT with threshold $\tau$. We derive an upper bound on the Type II error probability of such tests expressed in terms of the R\'enyi divergence of order $\lambda\in(0,1)$. This result is formalized in the following theorem, whose proof is deferred to Appendix \ref{app:thm_2}. 
\begin{theorem}\label{theorem:beta_upper_bound_renyi}
    Let $P_1^{n}$ be absolutely continuous with respect to $P_0^{n}$. Let $\phi:\mathcal{X}^n \to \{0,1\}$ be an LLRT with threshold $\tau\in \mathbb{R}$, and let $\alpha_\phi$ and $\beta_\phi$ denote the Type I error and Type II error probabilities of $\phi$, respectively. Then, one has
    \begin{equation}\label{eq:beta_upper_bound_renyi}
        \beta_\phi < \inf_{\lambda\in(0,1)} \frac{e^{(\lambda-1)D_{\lambda}(P_1^{n} \| P_0^{n})} - \alpha_\phi e^{\lambda \tau}}{e^{(\lambda-1)\tau}}.
    \end{equation}
\end{theorem}

\section{Relationship between Type I and II error probabilities}\label{sec:phase_transition}

In this section, we investigate the regime where the Type I error is constrained to decay exponentially with a rate $c\in \mathbb{R^+}$, i.e., $\alpha_n \leq e^{-nc}$. We provide a finite-sample characterization of the Type II error probability, recovering the sharp phase transition threshold identified by the divergence  $D(P_1\|P_0)$.

We formalize this result in the following theorem, whose proof is deferred to Appendix \ref{app:thm_3}.

\begin{theorem}\label{th:phase_transition}
    Let $\alpha_n\leq e^{-nc}$ be the constraint on the Type I error probability. 
    Then, the following property holds for the Type II error probability:

    \begin{equation}
         \lim_{n\to \infty} \beta_n\left(e^{-nc}\right) = \begin{cases}
             1 &\mbox{if } c>D(P_1\| P_0),\\
             0 &\mbox{if } c<D(P_1\| P_0).
         \end{cases}
    \end{equation}
    In particular, for any finite $n$, if $c>D(P_1\| P_0)$ it holds
    \remove{
    \begin{equation}\label{eq:c>}
        \beta_n(e^{-nc})\geq 1-e^{-n\sup_{\lambda>1}\left\{\frac{\lambda-1}{\lambda}\big(c-D_{\lambda}(P_1\|P_0)\big)\right\}},
    \end{equation}
    whereas, if $c<D(P_1\| P_0)$ it holds
    \begin{equation}\label{eq:c<}
        \beta_n(e^{-nc})<e^{-n\sup_{\lambda\in(0,1)}\left\{\frac{1-\lambda}{\lambda}\big(D_{\lambda}(P_1\|P_0)-c\big)\right\}}.
    \end{equation}
    }
        \begin{equation}\label{eq:c>}
        \beta_n(e^{-nc})\geq 1-\inf_{\lambda>1}\left(e^{-n(c-D_{\lambda}(P_1\|P_0))}\right)^{\frac{\lambda-1}{\lambda}},
        \end{equation}
        whereas, if $c<D(P_1\| P_0)$ it holds
        \begin{equation}\label{eq:c<}
            \beta_n(e^{-nc})<\inf_{\lambda\in(0,1)}e^{-\frac{1-\lambda}{\lambda}n(D_{\lambda}(P_1\|P_0)-c)}.
        \end{equation}
\end{theorem}

We remark that for \eqref{eq:c>},  since $c>D(P_1\|P_0)$, the non-decreasing property of $\lambda \to D_{\lambda}(P_1\|P_0)$ for $\lambda\in(0,+\infty)$ and the convergence $\lim_{\lambda\to 1^+} D_{\lambda}(P_1\|P_0) = D(P_1\|P_0)$ guarantee the existence of a $\lambda>1$ such that $c-D_{\lambda}(P_1\|P_0)>0$. Analogously, for \eqref{eq:c<}, since $c<D(P_1\|P_0)$,  $\lim_{\lambda\to 1^-} D_{\lambda}(P_1\|P_0) = D(P_1\|P_0)$ implies that there exists a $\lambda\in(0,1)$ such that  $D_{\lambda}(P_1\|P_0)-c>0$.

Theorem \ref{th:phase_transition} provides a finite-sample counterpart to the asymptotic results of Blahut \cite{Blahut} and Han and Kobayashi \cite{Han_strongconverse}. While their analysis revealed that the  divergence $D(P_1\|P_0)$ acts as a sharp phase transition threshold separating feasible and infeasible regimes as $n\to \infty$, our result explicitly quantifies the rate of convergence for finite $n$. In particular, we demonstrate that the Type II error converges to 1 exponentially fast when the Type I error decays at a rate larger than $D(P_1\|P_0)$, and conversely vanishes exponentially fast when the Type I error decays at a rate slower than $D(P_1\|P_0)$.

Our result also complements the classical Chernoff-Stein Lemma and its recent refinements by addressing a more stringent error regime. The Chernoff–Stein Lemma characterizes the optimal exponential decay rate of one error type under the constraint that the other error is bounded by a constant. Specifically, if we require the Type II error to be bounded by a fixed constant $\delta\in(0,1)$, the lemma guarantees that the Type I error can decay \emph{asymptotically} with a rate equal to $D(P_1\|P_0)$. Espinosa \textit{et al.} \cite{Espinosa} strengthened this by showing that one can maintain the optimal rate $D(P_1\|P_0)$  for the Type I error even under a sub-exponential decay for the Type II error. Our result demonstrates that if one is willing to ``back off" slightly from the optimal rate---accepting a Type I error rate $c < D(P_1\|P_0)$---the Type II error also can be made to decay exponentially fast, rather than sub-exponentially. Furthermore, given the non-asymptotic nature of our bounds, they enable the estimation of the sample size required to achieve specific error targets, as argued in Corollary \ref{cor:1}.

We conclude by observing that Theorem \ref{th:phase_transition} admits an equivalent symmetric formulation obtained by exchanging the roles of the errors. Specifically, if the Type II error probability is required to decay exponentially with rate $c$, then whenever $c>D(P_0\|P_1)$  the Type I error converges to 1 exponentially fast, whereas if $c< D(P_0\|P_1)$ the Type I error can be made to decay exponentially fast.

\section{Comparisons}\label{sec:plots}
Beyond characterizing the phase transition between feasible and infeasible regimes, and providing a strong converse result, our bound also yields significant improvements in the finite-sample regime. In this section, we show that our converse bound in Eq.~\eqref{eq:beta_lower_bound_renyi_additive} strictly improves upon existing results, yielding tighter lower bounds on the Type II error probability across a variety of settings. For this purpose, we present several numerical experiments
focusing on two classical testing scenarios. Note that our comparisons focus on the Chernoff-Stein regime, which is different from settings where both error types are required to be bounded (e.g.,~\cite{kazemi2025sample}).

\remove{
For this purpose, we begin by considering the simple case of hypothesis testing between two Gaussian distributions, for which we provide an analytic comparison. In particular, we focus on a ``stringent'' regime---corresponding either to a stringent Type~I error constraint (e.g., exponential decay) or to a limited sample size---and show that our bound accurately captures the behavior of the exact Type~II error probability $\beta_n(\varepsilon)$. We then complement this analysis with numerical experiments in two classical testing scenarios.
}

The first scenario considers testing between two Bernoulli distributions, where $P_0=\mathrm{Bern}(1/2)$ and $P_1=\mathrm{Bern(1/2+\delta)}$, with $\delta>0$.  
The second considers two Gaussians with $P_0=\mathcal{N}(\mu, 1)$ and $P_1=\mathcal{N}(\mu+\delta, 1)$, where the mean $\mu=2$ and $\delta>0$.

\begin{figure}
    \centering
    \setlength{\tabcolsep}{1pt} 
    
    \begin{subfigure}[b]{\columnwidth}
        \centering
        \includegraphics[width=\linewidth, trim={0 0.1cm 0 0}, clip]{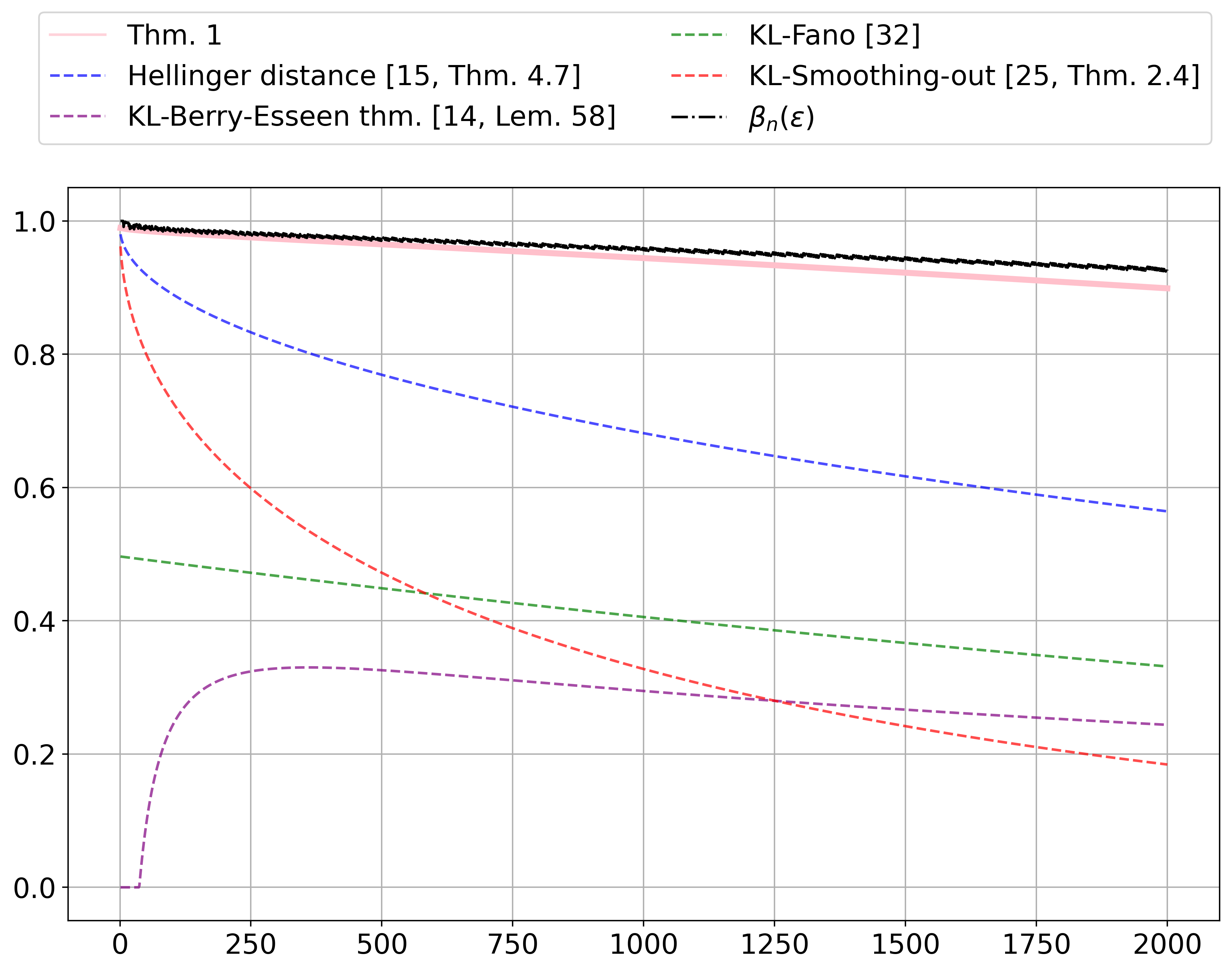}
        \caption*{Constant: $\varepsilon = 0.01$}
    \end{subfigure}

    \vspace{0.1cm} 
    
    \begin{subfigure}[b]{\columnwidth}
        \centering
        \includegraphics[width=\linewidth, trim={0 0.1cm 0 0}, clip]{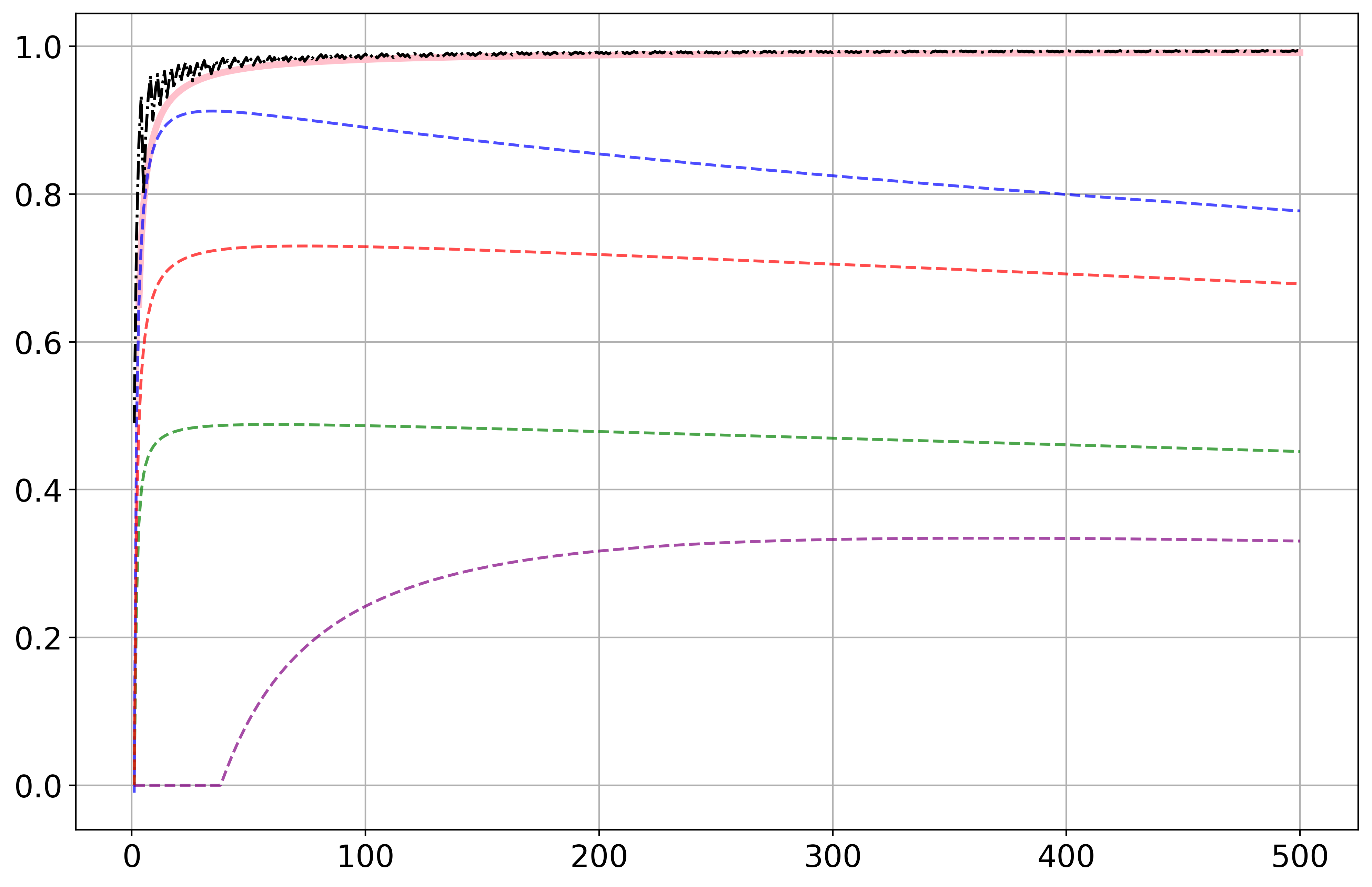}
        \caption*{Linear: $\varepsilon = 1/n$}
    \end{subfigure}
    
    \vspace{0.1cm}
    \begin{subfigure}[b]{\columnwidth}
        \centering
        \includegraphics[width=\linewidth, trim={0 0 0 0}, clip]{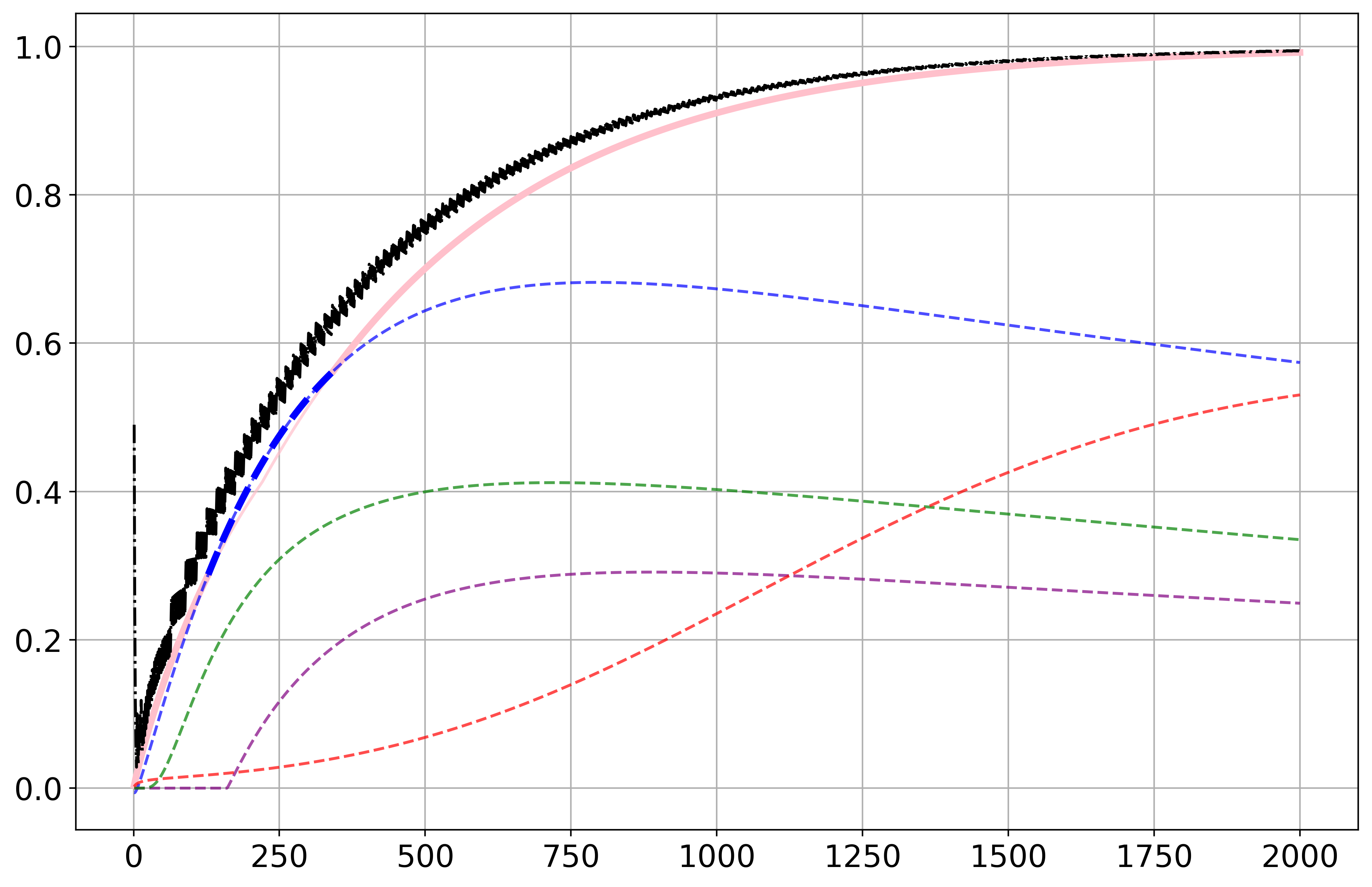}
        \caption*{Exponential: $\varepsilon = e^{-nc}, c=20\,D(P_1\|P_0)$}
    \end{subfigure}

    \caption{Bernoulli testing ($P_0=\text{Bern}(1/2)$ vs $P_1=\text{Bern}(1/2+\delta)$, with $\delta=0.01$) under three Type I error ($\varepsilon$) regimes: constant (top), linear (middle) and exponential (bottom). The $x$-axis denotes the sample size $n$, and the $y$-axis the Type II error.} 
    \label{fig:bernoulli}
\end{figure}

\begin{figure}
    \centering
    \setlength{\tabcolsep}{1pt} 
    
    \begin{subfigure}[b]{\columnwidth}
        \centering
        \includegraphics[width=\linewidth, trim={0 0.1cm 0 0}, clip]{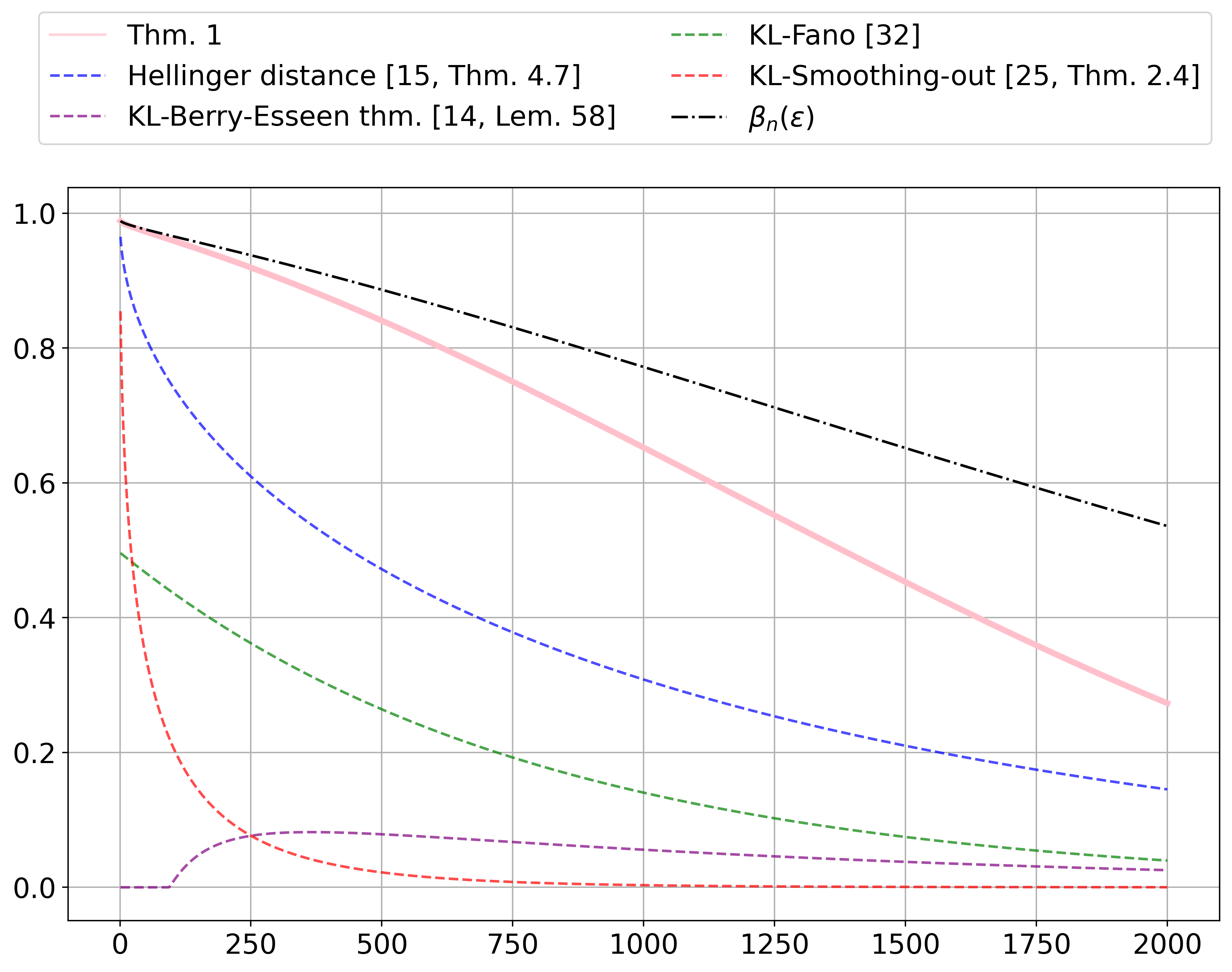}
        \caption*{Constant: $\varepsilon = 0.01$}
    \end{subfigure}
    
    \vspace{0.1cm} 
    
    \begin{subfigure}[b]{\columnwidth}
        \centering
        \includegraphics[width=\linewidth, trim={0 0.1cm 0 0}, clip]{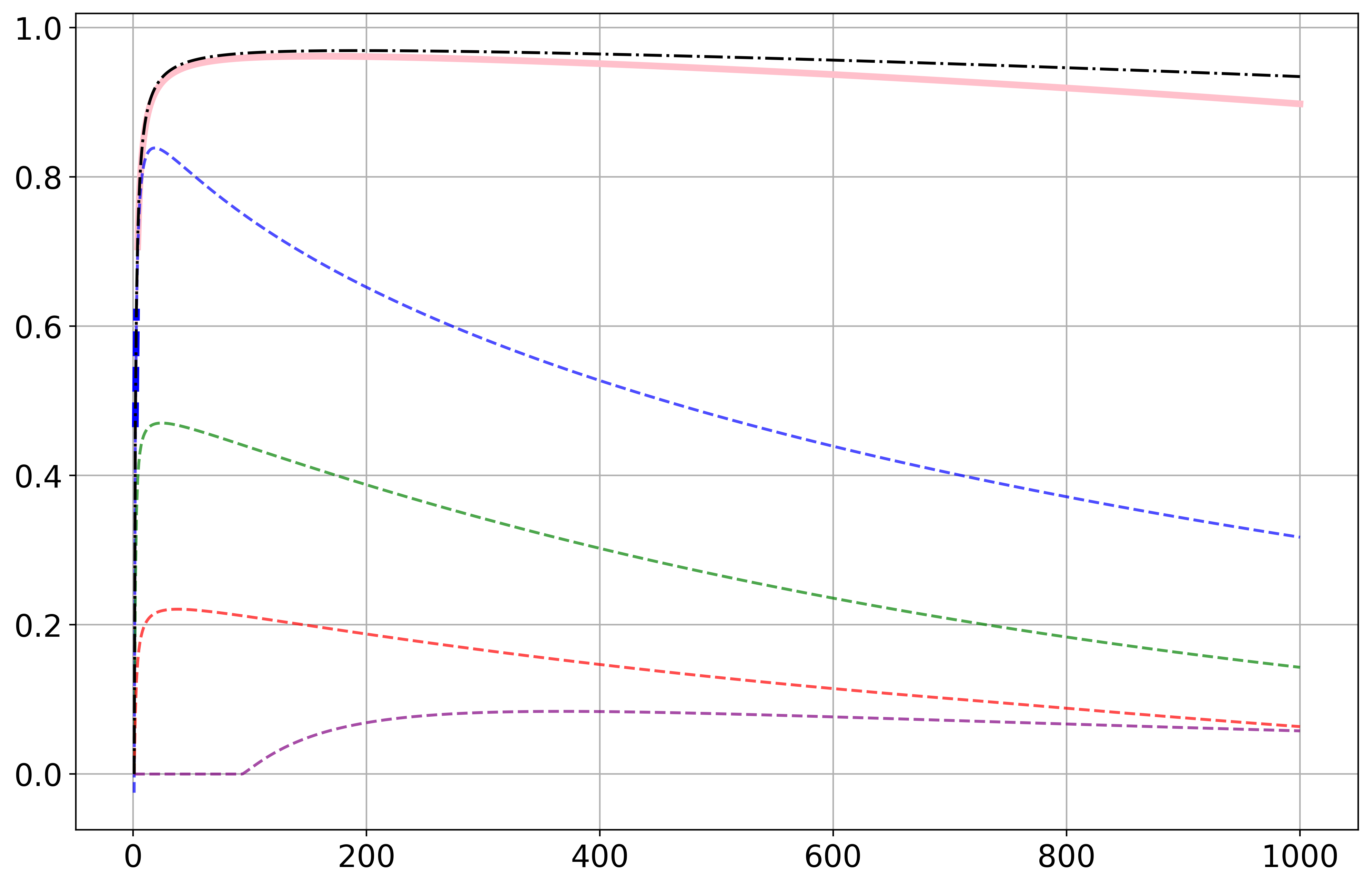}
        \caption*{Linear: $\varepsilon = 1/n$}
    \end{subfigure}
    
    \vspace{0.1cm}
    
    \begin{subfigure}[b]{\columnwidth}
        \centering
        \includegraphics[width=\linewidth, trim={0 0 0 0}, clip]{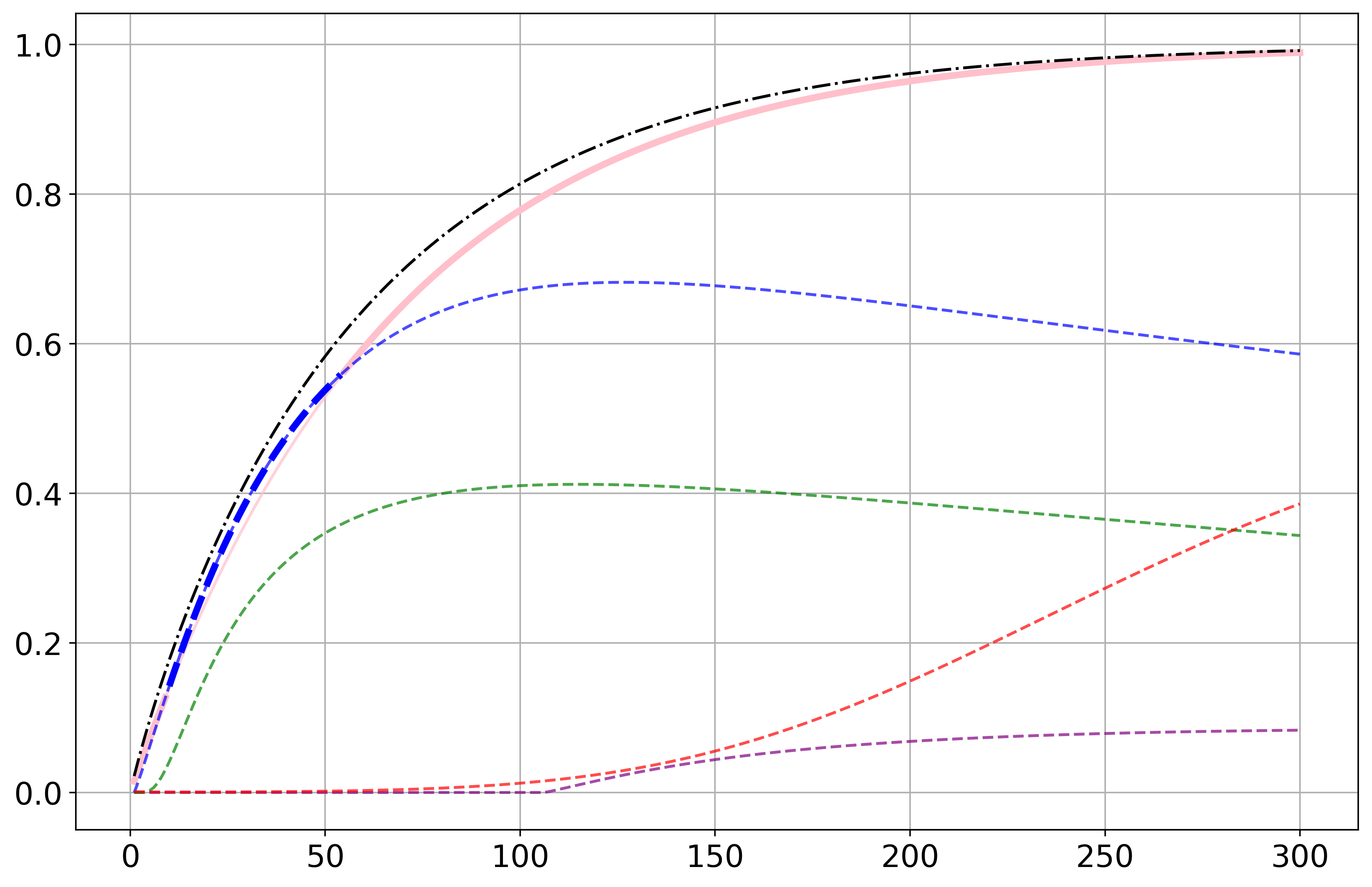}
        \caption*{Exponential: $\varepsilon = e^{-nc}, c=20\,D(P_1\|P_0)$}
    \end{subfigure}
    
    \caption{Gaussian testing ($P_0=\mathcal{N}(2,1)$ vs $P_1=\mathcal{N}(2+\delta,1)$, with $\delta=0.05$) under three Type I error ($\varepsilon$) regimes: constant (top), linear (middle) and exponential (bottom). The $x$-axis denotes the sample size $n$, and the $y$-axis the Type II error.}\label{fig:gaussian}
\end{figure}

For both scenarios, we benchmark our converse bound, under various Type I error constraints, against several well-known lower bounds. Three are based on the KL divergence, which includes Fano's inequality~\cite{fano1961transmission}, a bound derived from the Berry-Esseen theorem~\cite[Lemma 58]{polyanskiy2010channel}, and one from the smoothing-out method presented in \cite[Thm 2.4]{liu2020second}. The other lower bound we compare against is based on the Hellinger distance \cite[Thm 4.7]{bar2002complexity}. We exclude the bound obtained through the blowing-up lemma since it is strictly dominated by the smoothing-out bound in the regimes considered. In addition, for each setting, we compute the exact Type II error probability $\beta_n(\varepsilon)$ via the Neyman-Pearson Lemma. Figures \ref{fig:bernoulli} and \ref{fig:gaussian} illustrate these comparisons under three regimes for the constraint on the Type II error (constant, linear and exponential decay). 
For the Gaussian example, explicit computations are provided in Appendix~\ref{app:analytical_derivation_gaussians} to compare the behavior of the various bounds. In particular, the real Type II error behave as
\begin{equation}
     \beta_n(e^{-nc}) \sim 1 - \exp\left( -n \left(\sqrt{c} - \sqrt{D(P_1\|P_0)}\right)^2 \right), \label{eq:errorGaussian}
\end{equation}
with $c>D(P_1\|P_0)$, while Eq.~\eqref{eq:c>} with $\lambda=\sqrt{c/D(P_1\|P_0)}>1$ reduces to 
\begin{equation}
    \beta_n(e^{-nc}) \geq 1 - \exp\left( -n \left(\sqrt{c} - \sqrt{D(P_1 \| P_0)}\right)^2 \right),
\end{equation}
matching the exact behaviour from Eq.~\eqref{eq:errorGaussian}. By contrast, the converse bounds existing in the literature that leverage the Kullback-Leibler divergence or Hellinger distance fail to capture the convergence to $1$ of $\beta_n$. Additional scenarios of comparison are considered in Appendix \ref{app:comparisons}.

\bibliographystyle{IEEEtran}
\bibliography{references}
\newpage
\onecolumn
\appendices

\section{Proof of Theorem \ref{theorem:beta_lower_bound_renyi_additive}}\label{app:thm_1}
    We remark that the lower bound can be established via both the variational representation of the R\'enyi divergence and the Data-Processing Inequality (DPI). In the analysis that follows, we present the proof utilizing the DPI.

    Let $\mathrm{Bern}(p)$ denote the Bernoulli distribution with parameter $p$, and let $A$ be the acceptance region of an optimal decision rule that achieves the minimum Type II error in (\ref{eq:Beta_eps_def}). By applying the DPI on $D_\lambda(P_1^{n}\| P_0^{n})$ with $\lambda>1$, one gets
    \begin{equation}\label{eq:DPI_application}
        D_{\lambda}(P_1^{n}\| P_0^{n})) \geq D_{\lambda}(\mathrm{Bern}(P_1^{n}(A))\| \mathrm{Bern}(P_0^{n}(A))).
    \end{equation}
    From (\ref{eq:DPI_application}) by multiplying both sides by $\lambda-1$ and exponentiating, we obtain
    \begin{align}
        e^{(\lambda-1)D_{\lambda}(P_1^{n}\| P_0^{n}))}&\geq P_1^{n}(A)^\lambda P_0^{n}(A)^{1-\lambda} + P_1^{n}(A^c)^\lambda P_0^{n}(A^c)^{1-\lambda}\nonumber\\
        &=(1-\beta_n(\varepsilon))^\lambda \alpha_n^{1-\lambda}+ \beta_n(\varepsilon)^{\lambda}(1-\alpha_n)^{1-\lambda}\nonumber\\
        &=(1-\beta_n(\varepsilon))^\lambda \left(\frac{1}{\alpha_n}\right)^{\lambda-1}+ \beta_n(\varepsilon)^{\lambda}\left(\frac{1}{1-\alpha_n}\right)^{\lambda-1}\nonumber\\
        &\geq (1-\beta_n(\varepsilon))^\lambda \left(\frac{1}{\alpha_n}\right)^{\lambda-1}.\label{eq:DPI_step_1}
    \end{align}
    From (\ref{eq:DPI_step_1}), expressing in terms of $\beta_n(\varepsilon)$, it follows that
    \begin{align}\label{eq:lb_for_fixed_lambda_1}
        \beta_n(\varepsilon)\geq& 1-\left(\alpha_ne^{D_{\lambda}(P_1^{n}\| P_0^{n}))} \right)^{\frac{\lambda-1}{\lambda}}.
    \end{align}
    From (\ref{eq:lb_for_fixed_lambda_1}), since $\alpha_n\leq \varepsilon$ and the function $x\to x^{\frac{\lambda-1}{\lambda}}$ is increasing for $\lambda>1$, we get
    \begin{equation}\label{eq:lb_for_fixed_lambda_2}
        \beta_n(\varepsilon)\geq1-\left(\varepsilon e^{D_{\lambda}(P_1^{n}\| P_0^{n}))} \right)^{\frac{\lambda-1}{\lambda}}.
    \end{equation}
    Since (\ref{eq:lb_for_fixed_lambda_2}) holds for any $\lambda>1$, we obtain (\ref{eq:beta_lower_bound_renyi_additive}).

    \smallskip
    Similarly, by starting from $D_\lambda(P_0^{n}\| P_1^{n})$ rather than $D_\lambda(P_1^{n}\| P_0^{n})$, we obtain 
    \begin{equation*}
        \beta_n(\varepsilon)\geq \sup_{\lambda>1}\left((1-\varepsilon)^{\frac{\lambda}{\lambda-1}}\,e^{-D_\lambda(P_0^{n}\| P_1^{n})}\right).
    \end{equation*}
 Finally, by exploiting the tensorization property of the R\'enyi divergence for i.i.d. observations, we get the following lower bound on $\beta_n(\varepsilon)$
 \begin{equation*} \beta_n(\varepsilon)\geq  \max\Bigg\{1-\inf_{\lambda>1}(\varepsilon\, e^{nD_\lambda(P_1\| P_0)})^{\frac{\lambda-1}{\lambda}}, \quad\sup_{\lambda>1}\left((1-\varepsilon)^{\frac{\lambda}{\lambda-1}}\,e^{-nD_\lambda(P_0\| P_1)}\right)\Bigg\},
 \end{equation*}
concluding the proof.

\section{Proof of Corollary \ref{cor:1}}\label{app:cor}
    From (\ref{eq:beta_lower_bound_renyi_additive}) we have the following lower bound on $\beta_n(\varepsilon)$:
    \begin{align}\label{eq:lb_add_tensorization}
        \beta_n(\varepsilon)\geq&  \max\Bigg\{1-\inf_{\lambda>1}(\varepsilon\, e^{nD_\lambda(P_1\| P_0)})^{\frac{\lambda-1}{\lambda}}, \quad\sup_{\lambda>1}\left((1-\varepsilon)^{\frac{\lambda}{\lambda-1}}\,e^{-nD_\lambda(P_0\| P_1)}\right)\Bigg\}.
    \end{align}
    
    From (\ref{eq:lb_add_tensorization}), for a fixed $\lambda>1$, one can see that if the number of samples 
    \begin{equation*}
        n<\frac{\log(1/\varepsilon)}{D_\lambda(P_1\| P_0)},
    \end{equation*}
     the Type II error probability is bounded away from 0. More specifically, consider a sample size $n=\frac{\log(1/\varepsilon)-\log(\Delta^{\frac{\lambda}{\lambda-1}})}{D_\lambda(P_1\| P_0)}$, for some $\Delta>1$. From (\ref{eq:lb_add_tensorization}) one gets the lower bound 
    \begin{equation*}
        \beta_n(\varepsilon)\geq 1-\frac{1}{\Delta}.
    \end{equation*}
    Consequently, a necessary condition for the Type II error probability to vanish is that the sample size scales as $\Omega\left(\frac{\log(1/\varepsilon)}{D_\lambda(P_1\| P_0)}\right)$.

    Furthermore, if we require a test that achieves a Type II error probability below a target threshold $\delta$ (i.e., $\beta_n(\varepsilon)\leq \delta$), we can invert the bound in (\ref{eq:lb_add_tensorization}) for a fixed $\lambda$. This yields the following necessary condition on the sample size:
    \begin{align*}
         n\geq& \max\Bigg\{\frac{1}{D_{\lambda}(P_0\| P_1)}\left(\log \frac{1}{\delta}-\frac{\lambda}{\lambda-1}\log\frac{1}{1-\varepsilon}\right),\frac{1}{D_\lambda(P_1\|P_0)}\left(\log\frac{1}{\varepsilon}-\frac{\lambda}{\lambda-1}\log\frac{1}{1-\delta}\right)\Bigg\}.
    \end{align*}

\section{Proof of Theorem \ref{theorem:beta_upper_bound_renyi}}\label{app:thm_2}
    Let $A$ be the region of $\mathcal{X}^n $ in which the test $\phi$ accepts the alternative hypothesis, i.e., $\log \frac{p_1(\mathbf{x})}{p_0(\mathbf{x})}\geq \tau$ for all $x\in A$, where $p_1(\mathbf{x})$ and $p_0(\mathbf{x})$ are the probability density functions of $P_1^{n}$ and $P_0^{n}$ with respect to a dominating measure $\mu$.

    For an arbitrary $\lambda\in(0,1)$ the following holds:
    \begin{align}
        e&^{(\lambda-1)D_{\lambda}(P_1^{n} \| P_0^{n})}\nonumber\\ &= \int_{\mathbf{x}\in \mathcal{X}^n} \left(\frac{p_1(\mathbf{x})}{p_0(\mathbf{x})}\right)^{\lambda} p_0(\mathbf{x}) \,d\mu(\mathbf{x})\nonumber\\
        &=\int_{\mathbf{x}\in A} \left(\frac{p_1(\mathbf{x})}{p_0(\mathbf{x})}\right)^{\lambda} p_0(\mathbf{x}) \,d\mu(\mathbf{x})+ \int_{\mathbf{x}\in A^c} \left(\frac{p_1(\mathbf{x})}{p_0(\mathbf{x})}\right)^{\lambda} p_0(\mathbf{x}) \,d\mu(\mathbf{x})\nonumber\\
        &\geq \int_{\mathbf{x}\in A} e^{\lambda\tau} p_0(\mathbf{x}) \,d\mu(\mathbf{x})+ \int_{\mathbf{x}\in A^c} \left(\frac{p_1(\mathbf{x})}{p_0(\mathbf{x})}\right)^{\lambda} p_0(\mathbf{x}) \,d\mu(\mathbf{x})\nonumber\\&\quad\mbox{(since $p_1(\mathbf{x})/p_0(\mathbf{x})\geq e^\tau$ for all $\mathbf{x}\in A$)}\nonumber\\
        &=e^{\lambda\tau} P_0^{n}(A)+ \int_{\mathbf{x}\in A^c} \left(\frac{p_1(\mathbf{x})}{p_0(\mathbf{x})}\right)^{\lambda-1} \left(\frac{p_1(\mathbf{x})}{p_0(\mathbf{x})}\right)p_0(\mathbf{x}) \,d\mu(\mathbf{x})\nonumber\\
        &=e^{\lambda\tau} P_0^{n}(A)+ \int_{\mathbf{x}\in A^c} \left(\frac{p_0(\mathbf{x})}{p_1(\mathbf{x})}\right)^{1-\lambda} p_1(\mathbf{x}) \,d\mu(\mathbf{x})\nonumber\\
        &>e^{\lambda\tau} P_0^{n}(A) + \int_{\mathbf{x}\in A^c} e^{-(1-\lambda)\tau} p_1(\mathbf{x}) \,d\mu(\mathbf{x})\nonumber\\
        &\quad\mbox{(since $p_0(\mathbf{x})/p_1(\mathbf{x})>e^{-\tau}$ for all $\mathbf{x}\in A^c$)}\nonumber\\
        &= e^{\lambda\tau} P_0^{n}(A) + e^{-(1-\lambda)\tau}P_1^{n}(A^c)\nonumber\\
        &= e^{\lambda\tau} \alpha_\phi + e^{(\lambda-1)\tau} \beta_\phi.\label{eq:beta_upper_bound_step1}
    \end{align}
    Finally, optimizing (\ref{eq:beta_upper_bound_step1}) with respect to $\lambda$ yields  (\ref{eq:beta_upper_bound_renyi}), thereby concluding the proof.

\section{Proof of Theorem \ref{th:phase_transition}}\label{app:thm_3}

To prove the theorem, we first establish two auxiliary lemmas that relate the constant $c$ to the R\'enyi divergence $D_{\lambda}$.

\begin{lemma}\label{lemma:str_con}
    Let $c$ be a constant such that $c>D_{\lambda}(P_1\| P_0)$ for some $\lambda>1$. If a test satisfies the Type I error constraint, that is, $\alpha_n\leq e^{-nc}$, then its Type II error probability $\beta_n$ converges to 1, i.e.,
   \begin{equation}
       \lim_{n\to \infty} \beta_n = 1.
   \end{equation}
\end{lemma}
\begin{IEEEproof}
   Consider an arbitrary test that satisfies the Type I error constraint $\alpha_n\leq e^{-nc}$. Let $\delta>0$ be the strictly positive gap such that $c=D_{\lambda}(P_1\| P_0)+\delta$ for some $\lambda>1$.
   
   From Theorem~\ref{theorem:beta_lower_bound_renyi_additive}, we can bound its Type II error probability $\beta_n$ as follows: 
   \begin{align}
    \beta_n&\geq 1-\left(e^{-nc}e^{nD_{\lambda}(P_1\| P_0)}\right)^{\frac{\lambda-1}{\lambda}}\nonumber\\
    &= 1-\left(e^{-n(c-D_{\lambda}(P_1\| P_0))}\right)^{\frac{\lambda-1}{\lambda}}\nonumber\\
    &= 1-\left(e^{-n(D_{\lambda}(P_1\| P_0) +\delta-D_{\lambda}(P_1\| P_0))}\right)^{\frac{\lambda-1}{\lambda}}\nonumber\\
    &= 1-\left(e^{-n\delta}\right)^{\frac{\lambda-1}{\lambda}}=1-\left(e^{-n(c-D_{\lambda}(P_1\|P_0))}\right)^{\frac{\lambda-1}{\lambda}}.\label{eq:str_converse}
\end{align}
From (\ref{eq:str_converse}) since $\delta>0$ and $\lambda>1$, the subtracted term $\left(e^{-n\delta}\right)^{\frac{\lambda-1}{\lambda}}$ vanishes as $n\to\infty$. This implies that $\lim_{n\to\infty} \beta_n=1$. Since this holds for any test satisfying the constraint, the proof is complete.
\end{IEEEproof}

\begin{lemma}\label{lemma:ac}
    Let $c$ be a constant such that $c<D_{\lambda}(P_1 \| P_0)$ for some $\lambda\in(0,1)$. There exists a test that satisfies the Type I error constraint $\alpha_n\leq e^{-nc}$, and whose Type II error probability $\beta_n$ vanishes, i.e.,
    \begin{equation}
       \lim_{n\to \infty} \beta_n = 0.
   \end{equation}
\end{lemma}
\begin{IEEEproof}
    Let $\delta>0$ be the negative gap such that $c=D_{\lambda}(P_1\| P_0) -\delta$ for some $\lambda\in(0,1)$. We need to show that there exists a test that satisfies the constraint on the Type I error while ensuring that the Type II error vanishes as $n$ increases. For such a purpose, consider an LLRT $\phi$ with a threshold $\tau$ defined as follows
    \begin{equation}\label{eq:threshold}
        \tau=nD_{\lambda}(P_1\| P_0)-n\frac{\delta}{\lambda}.
    \end{equation}
    First, we verify that $\phi$ satisfies the Type I error constraint, i.e., its Type I error $\alpha_n\leq e^{-nc}$. Markov's inequality yields
    \begin{align*}
    \alpha_n =& P_0^{n}\left(\log \frac{p_1(\mathbf{x)}}{p_0(\mathbf{x})} \geq \tau\right)\\
    \leq& \frac{\mathbb{E}\left[\left(\frac{p_1(\mathbf{x)}}{p_0(\mathbf{x})}\right)^\lambda\right]}{e^{\lambda\tau}}\\
    =& e^{(\lambda-1)nD_{\lambda}(P_1\| P_0) - \lambda\tau}\\
    =& e^{(\lambda-1)nD_{\lambda}(P_1\| P_0) - \lambda nD_{\lambda}(P_1\| P_0) +n\delta}\nonumber\\
    =& e^{-n(D_{\lambda}(P_1 \| P_0)-\delta)} = e^{-nc}.
\end{align*}
Thus, $\alpha_n\leq e^{-nc}$, satisfying the constraint. 

It remains to prove that the Type II error probability decays as $n$ increases. Specifically, we show that it decays exponentially as a function of $n$. From Theorem \ref{theorem:beta_upper_bound_renyi} the Type II error $\beta_n$ of $\phi$ satisfies
    \begin{align}
    \beta &< \frac{e^{(\lambda-1)nD_{\lambda}(P_1 \| P_0)} - \alpha_n e^{\lambda \tau}}{e^{(\lambda-1)\tau}}\nonumber\\
    &<\frac{e^{(\lambda-1)nD_{\lambda}(P_1 \| P_0)}}{e^{(\lambda-1)\tau}}\nonumber\\
    &=  e^{(\lambda-1)nD_{\lambda}(P_1 \| P_0)-(\lambda-1)\tau}\nonumber\\
    &= e^{(\lambda-1)nD_{\lambda}(P_1 \| P_0)-(\lambda-1)nD_{\lambda}(P_1\| P_0) +(\lambda-1)\frac{n\delta}{\lambda}}\nonumber\\
    &= e^{\frac{\lambda-1}{\lambda}n\delta} = e^{\frac{1-\lambda}{\lambda}(-n\delta)}= e^{\frac{1-\lambda}{\lambda}(-n(D_{\lambda}(P_1\|P_0)-c))}.\label{eq:ub_application}
    \end{align}
    
    Finally, since $\lambda\in(0,1)$ and $\delta>0$, the right-hand side of (\ref{eq:ub_application}) vanishes as $n\to \infty$. Specifically, it vanishes exponentially fast with a rate that depends on the gap $\delta=D_\lambda(P_1\|P_0)-c$, implying $\lim_{n\to\infty} \beta_n = 0$. Thus, this concludes the proof.
\end{IEEEproof}

Armed with the two previous lemmas, we now proceed to provide the proof of Theorem \ref{th:phase_transition}, characterizing the threshold that separates the impossible regime from the achievable one.
 
\begin{IEEEproof}[Proof of Theorem~\ref{th:phase_transition}]
    The proof relies on the property that $\lim_{\lambda\to1} D_{\lambda}(P_1\| P_0) = D(P_1\| P_0)$. We need to consider two cases.

    \smallskip
    \noindent
    \textbf{Case $c>D(P_1\| P_0)$}: Since $D_{\lambda}(P_1\| P_0)$ approaches  $D(P_1\| P_0)$ as $\lambda\to 1^+$, for any fixed constant $c>D(P_1\| P_0)$, there exists a $\lambda>1$ sufficiently close to 1 such that 
    $$
    c> D_{\lambda}(P_1\| P_0)>D(P_1\| P_0).
    $$
    Thus, since for this choice of $\lambda>1$, the constant $c$ satisfies the condition of Lemma \ref{lemma:str_con}, it follows that for any test satisfying the Type I error  constraint $\alpha_n \leq e^{-nc}$, the Type II error probability $\beta_n$ converges to 1, i.e.,  $\lim_{n\to \infty}\beta_n=1$.

    \medskip
    \noindent
    \textbf{Case $c<D(P_1\| P_0)$}: Similarly, since $D_{\lambda}(P_1\| P_0)$ converges to  $D(P_1\| P_0)$ as $\lambda\to 1^-$, for any fixed constant $c<D(P_1 \| P_0)$, there exists a $\lambda\in(0,1)$ sufficiently close to 1 such that 
    $$
    c<D_{\lambda}(P_1\| P_0)<D(P_1 \| P_0).
    $$
    For this choice of $\lambda\in(0,1)$, the constant $c$ satisfies the condition of Lemma \ref{lemma:ac}. Consequently, there exists a test satisfying the Type I error constraint $\alpha_n \leq e^{-nc}$ while ensuring that the Type II error probability $\beta_n$ vanishes, i.e., $\lim_{n\to \infty}\beta_n=0$.

    \medskip
    Combining these two cases, we deduce that $c=D(P_1 \| P_0)$ represents the critical threshold for the phase transition. Furthermore, from (\ref{eq:str_converse}) and (\ref{eq:ub_application}), by optimizing on $\lambda$, we obtain (\ref{eq:c>}) and (\ref{eq:c<}),
    completing the proof.
\end{IEEEproof}

\section{Analytical Derivations for testing between Gaussians}\label{app:analytical_derivation_gaussians}
In this section, we analyze the behavior of the Type II error probability for a binary hypothesis testing problem between two Gaussian distributions, $P_0 \sim \mathcal{N}(\mu, 1)$ and $P_1 \sim \mathcal{N}(\mu+\delta, 1)$. We specifically consider the regime where the Type I error probability is constrained to decay exponentially fast, i.e., $\alpha_n \leq e^{-nc}$, with $c > D(P_1\|P_0)$. 

\subsection{Exact Asymptotic Analysis}
By the Neyman-Pearson lemma, the optimal test is a Log-Likelihood Ratio Test (LLRT). The log-likelihood ratio is given by:
\begin{align*}
    \log \frac{dP_1^n}{dP_0^n}(X^n) &= \sum_{i=1}^n \left( -\frac{(X_i - \mu - \delta)^2}{2} + \frac{(X_i - \mu)^2}{2} \right) \\
    &= \delta \sum_{i=1}^n (X_i - \mu) - \frac{n\delta^2}{2}.
\end{align*}
The decision rule is equivalent to comparing the sample mean $\bar{X}_n$ to a threshold $\gamma_n$. Specifically, we reject $H_0$ if $\bar{X}_n \geq \gamma_n$.

The threshold $\gamma_n$ is determined by the Type I error constraint. Under $H_0$, $\bar{X}_n \sim \mathcal{N}(\mu, 1/n)$. Thus,
\begin{equation*}
    \alpha_n = Q\left( \sqrt{n}(\gamma_n - \mu) \right) \leq e^{-nc},
\end{equation*}
where $Q(x) = \int_x^\infty \frac{1}{\sqrt{2\pi}}e^{-t^2/2}dt$. Assuming that the equality holds, and utilizing the asymptotic approximation $\log Q(x) \approx -x^2/2$ for large $x$, we have:
\begin{equation} \label{eq:threshold_approx}
    -\frac{n(\gamma_n - \mu)^2}{2} \approx -nc \implies \gamma_n \approx \mu + \sqrt{2c}.
\end{equation}
The Type II error probability $\beta_n$ corresponds to the probability that $\bar{X}_n < \gamma_n$ under $H_1$, where $\bar{X}_n \sim \mathcal{N}(\mu+\delta, 1/n)$.
\begin{align}
    \beta_n &= P_1(\bar{X}_n < \gamma_n) \nonumber \\
    &= \Phi\left( \sqrt{n}(\gamma_n - (\mu + \delta)) \right) \nonumber \\
    &= \Phi\left( \sqrt{n}(\gamma_n - \mu - \delta) \right). \label{eq:beta_exact}
\end{align}
Substituting the approximation from \eqref{eq:threshold_approx} into \eqref{eq:beta_exact}, and using the fact that the KL divergence between the two distributions is given by $D(P_1 \| P_0) = \frac{\delta^2}{2}$.
\begin{equation*}
    \beta_n \approx \Phi\left( \sqrt{n}(\sqrt{2c} - \delta) \right) = \Phi\left( \sqrt{2n}(\sqrt{c} - \sqrt{D(P_1\|P_0)}) \right).
\end{equation*}
When $c > D(P_1\|P_0)$, the argument is positive. Thus, we analyze how close $\beta_n$ is to 1 by considering $1 - \beta_n = Q(\cdot)$:
    \begin{equation}\label{eq:exact_type2_gaussian}
        \beta_n \sim 1 - \exp\left( -n \left(\sqrt{c} - \sqrt{D(P_1\|P_0)}\right)^2 \right).
    \end{equation}

\subsection{Converse via Theorem~\ref{theorem:beta_lower_bound_renyi_additive}}
First, we compute the Rényi divergence of order $\lambda$:
\begin{equation} \label{eq:Renyi_Gauss}
    D_\lambda(P_1 \| P_0) = \lambda \frac{\delta^2}{2} = \lambda D(P_1 \| P_0).
\end{equation}
From Eq.~(9) in Theorem 3, when $\alpha_n\leq e^{-nc}$ with $c>D(P_1\|P_0$), for any $n$, the Type II error is lower bounded by:
\begin{equation*}
    \beta_n \geq 1 - \inf_{\lambda > 1} \exp\left( -n \frac{\lambda-1}{\lambda} (c - D_\lambda(P_1 \| P_0)) \right).
\end{equation*}
Substituting \eqref{eq:Renyi_Gauss}, the exponent term becomes:
\begin{equation*}
    E(\lambda) \triangleq \left(1 - \frac{1}{\lambda}\right)\big(c - \lambda D(P_1 \| P_0)\big).
\end{equation*}
To find the tightest bound, we maximize $E(\lambda)$ with respect to $\lambda > 1$. The derivative is $\frac{d}{d\lambda} E(\lambda) = -D(P_1 \| P_0) + \frac{c}{\lambda^2}$ and setting it to zero yields $\lambda^* = \sqrt{c/D(P_1 \| P_0)}$. Since $c > D(P_1 \| P_0)$, $\lambda^* > 1$, which is valid. Evaluating the exponent at $\lambda^*$ gives $E(\lambda^*) = (\sqrt{c} - \sqrt{D})^2$ so that the bound becomes
\begin{equation*}
    \beta_n \geq 1 - \exp\left( -n \left(\sqrt{c} - \sqrt{D(P_1 \| P_0)}\right)^2 \right).
\end{equation*}
We observe that this matches the exact behavior of the Type~II error computed in Eq.~\eqref{eq:exact_type2_gaussian}.

\subsection{Converse via Fano's Inequality}
In this case, the lower bound on the Type II error $\beta_n$ is defined as follows:
\begin{equation*}
    \beta_n\geq\frac{e^{-nD(P_0\|P_1)-\log2}}{1-\alpha_n}.
\end{equation*}
Since $\alpha_n\leq e^{-nc}$, we obtain:
\begin{equation}\label{eq:fano}
     \beta_n\geq\frac{e^{-nD(P_0\|P_1)-\log2}}{1-e^{-nc}}.
\end{equation}
Since $c>D(P_1\|P_0)$, the denominator in \eqref{eq:fano} approaches 1 as $n \to \infty$. Consequently, the bound asserts: 
\begin{equation*} 
\beta_n \gtrsim e^{-n D(P_0\|P_1)}. 
\end{equation*}
However, from Eq.~\eqref{eq:exact_type2_gaussian} for $c > D(P_1\|P_0)$, the Type II error probability $\beta_n$ converges to 1. Thus, the inequality derived via Fano's Inequality is valid but provides no information about the rate at which $\beta_n$ approaches 1 in this setting.

\subsection{Converse via Hellinger Distance}
Let $H^2(P_0, P_1)$ denote the squared Hellinger distance between $P_0$ and $P_1$, defined as follows:
\begin{align*}
        H^2(P_0, P_1) &=1 - \int_{\mathcal{X}} \sqrt{p_0(\mathbf{x}) p_1(\mathbf{x})} \, d\mu(\mathbf{x}).
\end{align*}
From the approach presented in \cite[Thm 4.7]{bar2002complexity}, one derive the following converse bound on the Type II error:
\begin{equation*}
\beta_n\geq 1-\sqrt{1-(1-H^2(P_0,P_1))^{2n}}-\alpha_n.
\end{equation*}
Recalling that $1 - H^2(P_0, P_1) = e^{-\frac{\delta^2}{8}}$ and $D(P_1\|P_0)=\frac{\delta^2}{2}$, we have that $1 - H^2(P_0, P_1) = e^{-\frac{D(P_1\|P_0)}{4}}$. Moreover, since $\alpha_n\leq e^{-nc}$, we obtain:
\begin{equation}\label{eq:hellinger_bound}
\beta_n\geq 1-\sqrt{1-e^{-\frac{nD(P_1\|P_0)}{2}}}-e^{-nc}.
\end{equation}
For $c>D(P_1\|P_0)$, one can see that both the term $e^{-\frac{nD(P_1\|P_0)}{2}}$ and $e^{-nc}$ in Eq.~\eqref{eq:hellinger_bound} vanishes as $n\to\infty$. Therefore, as Fano's Inequality, it provides no information about the rate at which $\beta_n$ approaches 1.

\subsection{Converse via Berry-Esseen Theorem}
From ~\cite[Lemma 58]{polyanskiy2010channel}, one has the following lower bound on the Type II error:
\begin{equation*}
    \log(\beta_n)\geq -nD(P_0\|P_1) -\sqrt{nV_n}Q^{-1}\left(1-\alpha_n-\frac{B_n+\Delta}{\sqrt{n}}\right) + \log \Delta -\frac{1}{2}\log n,
\end{equation*}
where $V_n = \text{Var}_{P_0} \left[ \log \frac{p_0(\mathbf{X})}{p_1(\mathbf{X})} \right]$, $T_n = \mathbb{E}_{P_0} \left[ \left| \log \frac{p_0(\mathbf{X})}{p_1(\mathbf{X})} - D(P_0 \| P_1) \right|^3 \right]$, $B_n = 6 \frac{T_n}{V_n^{3/2}}$ and $\Delta>0$.

For the Gaussian case $P_0 \sim \mathcal{N}(\mu, 1)$ and $P_1 \sim \mathcal{N}(\mu+\delta, 1)$, we have that $V_n=\delta^2$, $T_n=|\delta|^3\sqrt{8/\pi}$ and $B_n=6\sqrt{8/\pi}$. Thus, by fixing $\Delta>0$ we obtain:
\begin{equation*}
    \log(\beta_n)\geq -nD(P_0\|P_1) -\sqrt{n\delta^2}Q^{-1}\left(1-\alpha_n-\frac{6\sqrt{8/\pi}+\Delta}{\sqrt{n}}\right) + \log \Delta -\frac{1}{2}\log n.
\end{equation*}
As $n\to \infty$, since $\alpha_n\leq e^{-nc}$, the argument of $Q^{-1}(\cdot)$ approaches 1. Using the approximation $Q^{-1}(1-\epsilon)\sim -\sqrt{2\log (1/\epsilon)}$, with $\epsilon=\alpha_n+\frac{6\sqrt{8/\pi}+\Delta}{\sqrt{n}}$, we get
\begin{equation}\label{eq:esseen_bound}
    \log(\beta_n)\geq -nD(P_0\|P_1) +\sqrt{n\delta^2}\sqrt{\log n-2\log(\sqrt{n}\alpha_n+6\sqrt{8/\pi}+\Delta)} + \log \Delta -\frac{1}{2}\log n.
\end{equation}
As $n\to \infty$, since $\alpha_n\leq e^{-nc}$ for $c>D(P_1\|P_0)$, the second term in \eqref{eq:esseen_bound} scales as $\sqrt{\delta^2n\log n}$. Thus, recalling that $D(P_1\|P_0)=\delta^2/2$, it follows that the term $-nD(P_0\|P_1)$ dominates the expression \eqref{eq:esseen_bound}. This implies that for $c>D(P_1\|P_0)$, the lower bound decays to 0. In contrast, the exact Type II error actually converges to 1 as shown in \eqref{eq:exact_type2_gaussian}.

\subsection{Converse via Smoothing Out}
Applying the smoothing-out method presented in \cite[Thm 2.4]{liu2020second} to this Gaussian setting, we have that for any $t>0$ the following bound holds:
\begin{equation*}
    \log(\beta_n)\geq -nD(P_0\|P_1)+\frac{1}{1-e^{-2t}}\log(1-\alpha_n)-nt-\frac{\delta^2}{2}\left(e^t-1\right)^2 -n\big(\cosh(2t)-1\big),
\end{equation*}
where $\cosh(t) = \frac{e^t + e^{-t}}{2}$.

To find the tightest bound, we have to observe how the parameter $t$ behaves. If $t$ is fixed, as $n\to \infty$, since $\alpha_n\leq e^{-nc}$, the terms $-nt$ and $-n(\cosh(2t)-1)$ dominate the lower bound, forcing it to go to $-\infty$ (i.e., $\beta_n\geq 0$). Thus, the parameter $t$ must vanish as $n$ increases. Using the Taylor expansions $e^{-2t}\approx1-2t$, $\log(1-x)\approx-x$ for small $x$, and $\cosh(2t)-1\approx2t^2$, the bound simplifies to:
\begin{equation*}
    \log(\beta_n) \gtrsim -nD(P_0\|P_1) - \frac{\alpha_n}{2t} - nt - 2nt^2.
\end{equation*}
Optimizing for $t$ yields $t\approx\sqrt{\alpha_n/2n}$. However, because $\alpha_n\leq e^{-nc}$, this choice makes the term $-nD(P_0\|P_1)$ the dominant one, as $n\to\infty$. Consequently, in the regime $c>D(P_1\|P_0)$, the lower bound fails to capture the fact that the exact Type II error $\beta_n$ converges to 1.

\section{Additional Comparisons}\label{app:comparisons}

\begin{figure*}[ht] 
    \centering
    
    \begin{subfigure}{\textwidth}
        \centering
        \includegraphics[width=\textwidth]{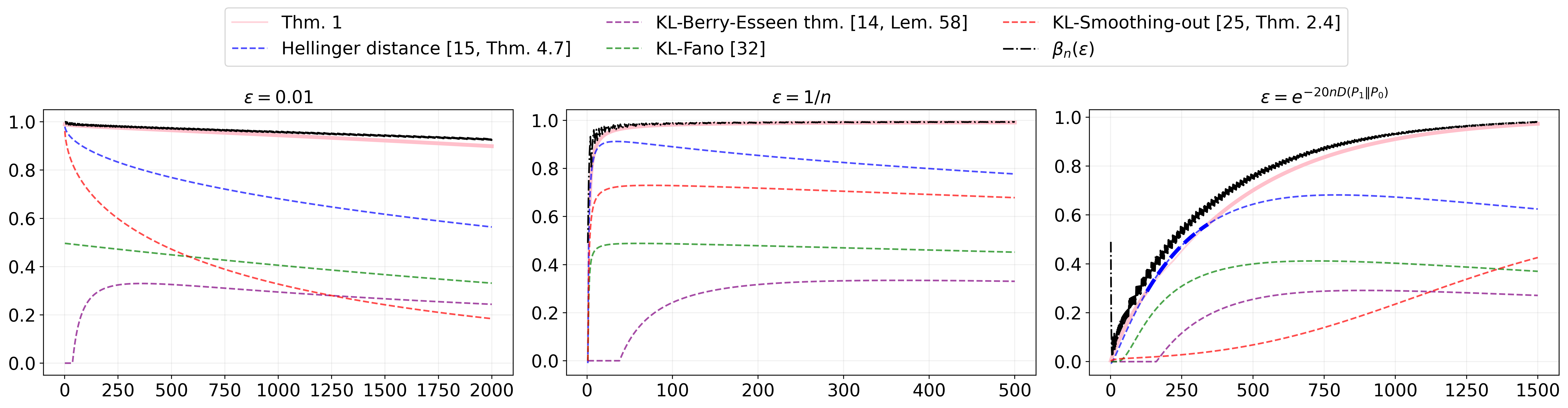}
        \label{fig:bern}
    \end{subfigure}
    
    \par 
    
    \begin{subfigure}{\textwidth}
        \centering
        \includegraphics[width=\textwidth]{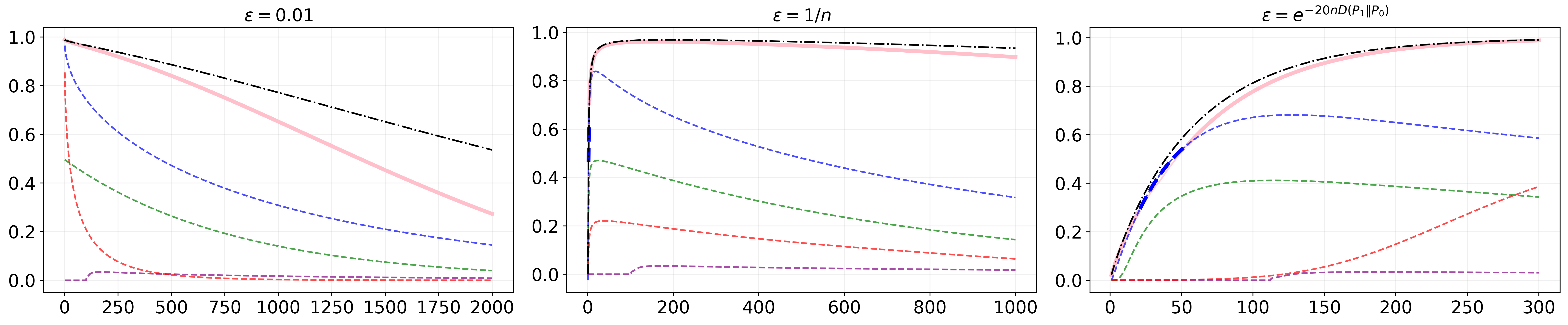}
        \label{fig:gauss}
    \end{subfigure}
    
    \caption{(\textbf{Top row}) Bernoulli testing ($P_0=\text{Bern}(1/2)$ vs $P_1=\text{Bern}(1/2+\delta)$, with $\delta=0.01$) under three Type I error ($\varepsilon$) regimes: constant (left), linear (middle), and exponential (right).
    (\textbf{Bottom row}) Gaussian testing ($P_0=\mathcal{N}(2,1)$ vs $P_0=\mathcal{N}(2+\delta,1)$, with $\delta=0.05$) under three Type I error ($\varepsilon$) regimes: constant (left), linear (middle), and exponential (right).
    The $x$-axis denotes the sample size $n$, and the $y$-axis the Type II error.}
    \label{fig:combined_testing}
\end{figure*}

\begin{figure*}[ht] 
    \centering
    
    \begin{subfigure}{\textwidth}
        \centering
        \includegraphics[width=\textwidth]{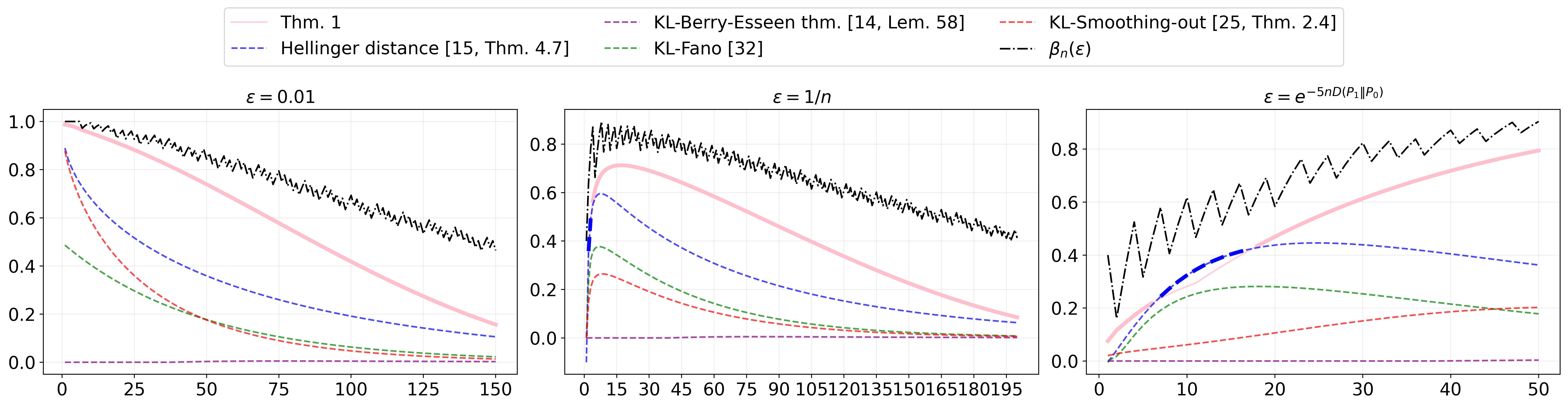}
    \end{subfigure}
    
    \par 
    
    \begin{subfigure}{\textwidth}
        \centering
        \includegraphics[width=\textwidth]{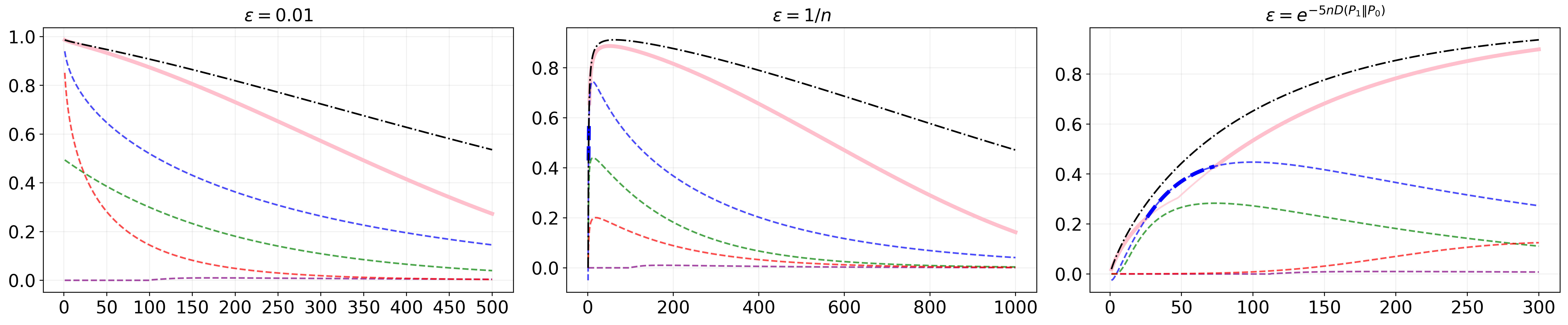}
    \end{subfigure}
    
    \caption{(\textbf{Top row}) Bernoulli testing ($P_0=\text{Bern}(1/2)$ vs $P_1=\text{Bern}(1/2+\delta)$, with $\delta=0.1$) under three Type I error ($\varepsilon$) regimes: constant (left), linear (middle), and exponential (right).
    (\textbf{Bottom row}) Gaussian testing ($P_0=\mathcal{N}(2,1)$ vs $P_0=\mathcal{N}(2+\delta,1)$, with $\delta=0.1$) under three Type I error ($\varepsilon$) regimes: constant (left), linear (middle), and exponential (right).
    The $x$-axis denotes the sample size $n$, and the $y$-axis the Type II error.}
\end{figure*}

\begin{figure*}[ht] 
    \centering
    
    \begin{subfigure}{\textwidth}
        \centering
        \includegraphics[width=\textwidth]{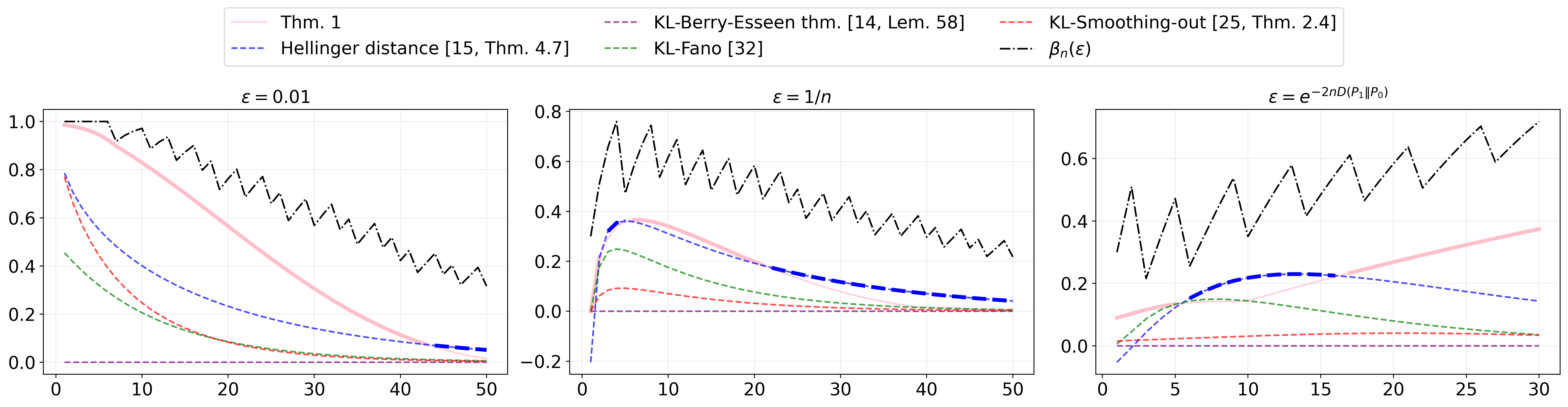}
    \end{subfigure}
    
    \par 
    
    \begin{subfigure}{\textwidth}
        \centering
        \includegraphics[width=\textwidth]{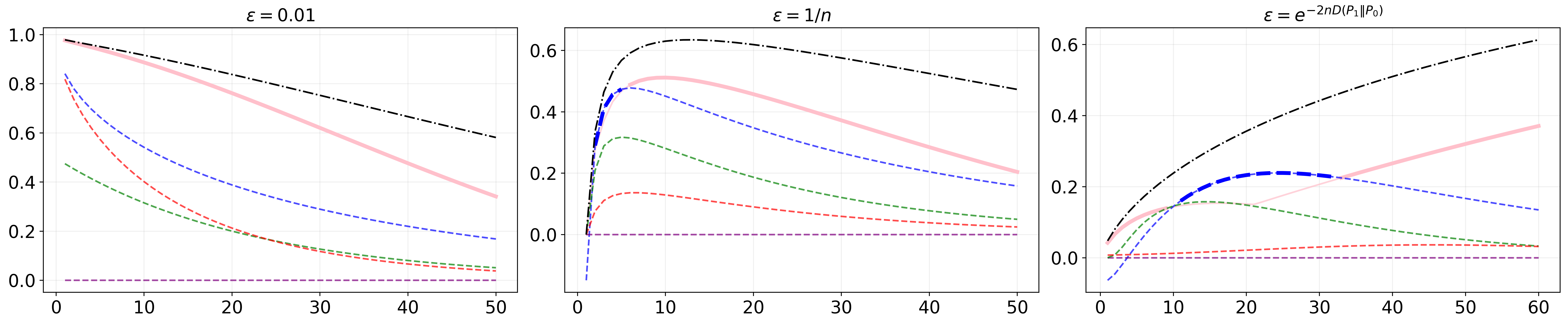}
    \end{subfigure}
    
    \caption{(\textbf{Top row}) Bernoulli testing ($P_0=\text{Bern}(1/2)$ vs $P_1=\text{Bern}(1/2+\delta)$, with $\delta=0.2$) under three Type I error ($\varepsilon$) regimes: constant (left), linear (middle), and exponential (right).
    (\textbf{Bottom row}) Gaussian testing ($P_0=\mathcal{N}(2,1)$ vs $P_0=\mathcal{N}(2+\delta,1)$, with $\delta=0.3$) under three Type I error ($\varepsilon$) regimes: constant (left), linear (middle), and exponential (right).
    The $x$-axis denotes the sample size $n$, and the $y$-axis the Type II error.}
\end{figure*}

\end{document}